\documentclass[12pt,preprint]{aastex}

\slugcomment{To appear in ApJL}

\shorttitle{The Masses of  Transition Circumstellar Disks}
\shortauthors{Cieza et al.}

\begin{document}

\title{The Masses of Transition Circumstellar Disks:  Observational Support for Photoevaporation Models}

\author{Lucas A. Cieza\altaffilmark{1,}\altaffilmark{2},
Jonathan J. Swift\altaffilmark{1},
Geoffrey S. Mathews\altaffilmark{1}
 \& Jonathan P. Williams\altaffilmark{1}
 }
 
\altaffiltext{1}{Institute for Astronomy, The University of Hawaii at Manoa, HI  96822}
\altaffiltext{2}{\emph{Spitzer} Fellow:  lcieza@ifa.hawaii.edu}

\begin{abstract}

We report deep Sub-Millimeter Array observations of 
26 pre-main-sequence (PMS) stars  with evolved  inner disks.
These observations measure the mass of the outer disk (r  $\sim$20-100 AU) across 
every stage of the dissipation of the inner disk (r $<$ 10 AU) as determined by the IR  spectral energy distributions (SEDs). 
We find that only  targets with  high mid-IR excesses are detected and have disk masses in the 1-5 M$_{Jup}$ range, 
while  most of our objects remain undetected to sensitivity levels  of  M$_{DISK}$ $\sim$0.2-1.5 M$_{Jup}$.   To put these 
results in a more general context, we collected publicly available data to construct the optical to millimeter  wavelength 
SEDs of over 120 additional PMS stars.  We find that the near-IR and mid-IR emission remain optically thick  in objects 
whose disk masses span  2 orders of magnitude ($\sim$0.5-50 $M_{Jup})$.  Taken together,  these results  imply that, 
\emph{in general,  inner disks start to dissipate only after the outer disk has been significantly depleted of mass.}  
This provides strong support for photoevaporation being one of the dominant processes driving disk evolution. 

\end{abstract}
\section{Introduction}

The vast majority of PMS stars in any given population are 
either accreting Classical T Tauri Stars (CTTSs)  with excess emission 
extending from the  near-IR to the millimeter or weak-line T Tauri Stars (WTTSs) 
with bare stellar photospheres. The fact that very few 
objects lacking near-IR excess show mid-IR or (sub)millimeter excess emission
implies that, once the inner disk dissipates,  the entire disk goes away very rapidly 
(Skrutskie et al. 1990; Wolk $\&$ Walter 1996; Cieza et al. 2007). 

To first order, the evolution of primordial disks is driven by viscous accretion. Viscous 
evolution models (Hartmann et al. 1998;  Hueso $\&$ Guillot 2005) are broadly consistent with the
observational constraints for disk masses, disk sizes,  and accretion rates as a function
of time;  however, they also predict smooth,  power-law  evolution of the disk properties.
This smooth evolution is inconsistent  with the very rapid disk dissipation 
that usually occurs after a much longer disk lifetime. Pure viscous evolution models also fail to 
explain the variety of SEDs of  the so called  ``transition  objects"  (the few objects 
that are caught in the short transition between typical CTTSs and bare stellar photospheres). 

Recent disk evolution models, known as ``UV-switch'' models, combine viscous evolution
with photoevaporation by the central star (Clarke et al. 2001; Alexander et al. 2006) and
are able to account for both the disk lifetimes of several million years and the short disk
dissipation timescales  ($\tau$ $<$ 0.5 Myr).  
According to these models, extreme ultraviolet  (EUV) photons originating in the 
stellar chromosphere, ionize and heat the circumstellar hydrogen. 
Beyond some critical radius, the thermal velocity of the ionized hydrogen exceeds its escape 
velocity and the material is lost in the form of a wind. At early stages,
the accretion rate dominates over the evaporation rate and the disk undergoes standard viscous 
evolution.  Later on, as the accretion rate drops,
the outer disk is no 
longer able to resupply the inner disk with material. At this point, the inner  disk drains on a viscous 
timescale and an inner hole is formed.
Once this inner hole has formed, EUV radiation very efficiently photoevaporates the inner edge of the 
disk and the disk rapidly dissipates from the inside out. 
Thus, the UV-switch model naturally accounts for disk 
lifetimes and dissipation timescales as well as for the SEDs of  some disks suggesting the presence 
of large inner holes. 
 
Photoevaporation, however, is not the only mechanism that has been proposed to explain
the opacity holes of  transition disks, some of which have now been confirmed by direct submillimeter  
imaging  (e.g., Brown  et al. 2008).  
Theoretical models of the dynamical interaction between 
forming planets and the disk (Lin $\&$ Papaloizous 1979,
Artymowicz $\&$ Lubow 1994) also predict the formation of inner holes
and gaps, and thus planet formation quickly became one of the
most exciting explanations proposed for the inner holes of
transition disks (Calvet et al. 2002; D'Alessio et al. 2005).
However, the planet formation process is not required to be far along in order
to affect the SED of a circumstellar disk.  Once primordial sub-micron
dust grains grow into somewhat larger bodies (r $\gg$ $\lambda$), most of the solid mass
ceases to interact with the radiation, and the opacity function
decreases dramatically. Dullemond $\&$ Dominik (2005) 
find that grain growth is a strong function of radius
it is more efficient in the inner 
regions where the surface density is higher and the dynamical timescales are 
shorter,  and hence can also produce opacity holes. 

Understanding  the processes operating in  transition disks is crucial 
for understanding disk evolution and planet formation. However, since multiple 
processes can result in similar  IR SEDs, additional observational 
constraints are necessary to establish their relative importance.
Here, we  report deep SMA observations of a sample 
of 26 PMS stars whose IR SEDs trace the dissipation of the inner 
disk (r$<$ 10 AU).  These observations provide information on  the  mass 
of  their  cold outer disks (r  $\sim$20-100 AU)  and help us to distinguish  
between different evolutionary scenarios. 
    
\section{Sample Selection and Observations}{\label{OBS}}

The stars in our sample were selected from the literature and meet the following 
criteria  {\it{i)}} are low-mass (A-type or later, but mostly K-M type stars) PMS stars 
with ages $\lesssim$ 10 Myrs,  {\it{ii)}} have 24 $\mu$m excesses,  
{\it{iii)}}  show evidence for  inner disk evolution. This evidence is in the 
form of decreased levels of near- and mid-IR excess (i.e. are in the  lower 
quartile of the SEDs of CTTSs presented 
by Furlan et al.  2006) and/or weak accretion (i.e. are WTTSs), and {\it{iv)}} 
are located within 140 pc of the sun. Following these criteria, we selected 11 objects
located in the Ophiuchus molecular cloud from Cieza et al. (2007),  10
objects from the Upper Scorpius association from Carpenter et al.  (2006), 
and  5 objects from the TW Hydra association from Low et al (2005).
Based on criteria {\it{ii)}} and {\it{iii)}}, our targets can be broadly 
classified as transition objects. However, we note that the precise 
definitions of what constitutes a transition object found in the literature 
are far from homogeneous. 

(Sub)millimeter interferometric observations of our 25 stars were 
conducted with the Submillimeter Array (SMA; Ho et al. 2004) during 
the summer of  2007 (July 5-31, compact configuration) and 
the spring of  2008 (April 24-May 17, compact-north configuration).  
Fifteen of our targets were observed with the 230 GHz/1300 $\mu$m 
receivers, 9 of them 
with the 345 GHz/ 850 $\mu$m receivers, and 
2 additional targets were observed at both frequencies. 
The upper and lower sideband data were used in both wavebands, 
resulting in a total bandwidth of 4\,GHz. 
Typical zenith opacities for our data were $\tau_{225\,{\rm GHz}}  = 0.08$--0.14. 
For each target source, the observations cycled between the target and
two gain calibrators (1037-295/1130-148 for TW Hydra objects, 1517-243/1733-130
for the Upper Sco Objects, and  1626-298/1733-130 for Ophiuchus objects), 
with 20-30 minutes on target and 4-7.5 minutes on each calibrator. 

The visibility data were calibrated with the MIR reduction
package\footnote{\tt  http://cfa-www.harvard.edu/$\sim$cqi/mircook.html}. 
The passband was flattened using $\sim$1 h  scans of  Uranus or 3c454.3 
and the complex gain solutions were obtained using the primary calibrators 
1037-295, 1517-243, and 1626-298.  These gains, applied to our secondary calibrators, 
served as a consistency  check for these solutions. 
Since none of the detected targets were resolved, their flux densities were 
measured by fitting a point source model to the visibility data. 
The upper limits were derived from the rms of the visibility amplitudes. The absolute 
flux scale was determined through observations of either Uranus,  Callisto,  or 
Neptune and is estimated to be accurate to 15$\%$. 

\section{Results: The Disk Masses of Transition Objects}{\label{RES}}

The 850 and 1300 $\mu$m SMA continuum measurements for 
our sample, along with 2.2 and 24 $\mu$m photometry from 2MASS and 
\emph{Spitzer} are presented in Table 1.  We detect only 6 out of the 
26 PMS stars in our sample:  ROXs 10B, ROXR1 29,  USco J160822.4-193004,
USco J160900.7-190852, TWA 3, and TWA 4. 
Andrews $\&$ Williams (2007)  show that disk masses obtained 
from modeling their IR and (sub)-mm SEDs are well described 
by a simple relation of the form M$_{DISK}$ = C$_{\nu}$$\times$ F$_{\nu}$,
where C$_{\nu}$ is a constant at a given frequency and F$_{\nu}$ is 
a (sub)millimeter flux at that same frequency. 
In  order to allow a direct comparison of our disk masses to relevant 
previous results
(see Sec.~\ref{SEDevo}), 
we   compute  C$_{\nu}$ at 230 and 345 GHz  from the ratios of derived  disk masses to observed 
fluxes  presented by Andrews $\&$ Williams (2005).  We adopt the mean 
values of these ratios for  the 33 Taurus stars with both 230 and 345 GHz 
data  for which they obtain disk masses from SED fitting. 
From the constants so derived, we obtain to  the following relations: 

\indent 1) M$_{DISK}$=8.0$\times$10$^{-5}$  [($\frac{F_\nu(\ 850)}{mJy}$)$\times$($\frac{d}{140  pc})^2$]   M$_\odot$ or  \\
\indent 2) M$_{DISK}$=1.7$\times$10$^{-4}$  [($\frac{F_\nu(1300)}{mJy}$)$\times$($\frac{d}{140  pc})^2$] M$_{\odot}$    

Using  these equations,  we find that the 6  targets detected  have disks masses in the 1-5 M$_{Jup}$ range, 
while most of the  undetected objects have 3-$\sigma$ upper limits in the 
$\sim$0.2-1.5 M$_{Jup}$ range (See Tab. 1). 
Since  Andrews $\&$ Williams (2007) find a median disk mass of 
5 M$_{Jup}$   for CTTSs in Taurus and Ophiuchus, 
we conclude that  the stars  in our sample, selected based on their inner 
disk properties, have disks significantly  less massive than those of 
typical CTTS stars.
We also find that the SMA targets with highest disk masses ($>$2 M$_{Jup}$),  
ROXs 10B, ROXR1 29, and USco 
USco-J160823.2-193001
 are also among those 
with the highest levels of IR excesses at IRAC wavelengths (See Fig. 1).   

\section{Discussion and Conclusions}{\label{DIS}}

\subsection{The SED  evolution of PMS Stars}{\label{SEDevo}}

In order to put the results of our SMA survey in a more general context 
of disk evolution,  we combined publicly available data to construct 
the optical, near-, mid-IR, and  (sub)millimeter wavelength SEDs of over 
120 additional PMS stars. We started from the (sub)millimeter targets 
in Taurus and Ophichus studied by Andrews $\&$ Williams 
(2005, 2007) and then searched for their  \emph{Spitzer} fluxes 
in the catalogs produced  by the ``Cores to Disks" (Evans et al. 2003) 
and Taurus (Padget et al. 2006) Legacy Projects
\footnote{\tt http://ssc.spitzer.caltech.edu/legacy/all.html}. 
For objects with both \emph{Spitzer} and (sub)millimeter data,
we collected the near-IR photometry from the 2MASS database and, when
available, the optical data from the literature. The complete SEDs of 
all these targets will be presented in a followup paper (Cieza et al., in prep). 
In this Letter, we focus on the analysis of the [K$_S$]-[24] colors of the 
extended sample (our SMA sample plus the objects discussed above) 
as a function of  disk mass. For consistency, we calculate all disk 
masses from Eq. 1 (or Eq. 2 if  850 $\mu$m data are 
not available), assuming the following distances:  50 pc for 
the TW Hydra association, 125 pc for Ophuichus, and 140 pc for 
the Taurus and Upper Sco  regions.    

In Fig.  2 we plot disk mass as a function of [K$_S$]-[24] 
color for the extended sample. 
This figure shows that our SMA study is significantly more sensitive than previous 
surveys.  One striking result from Figure 2 is that all the 
objects with disk masses larger  than $\sim$2 M$_{Jup}$ have  [K$_S$]-[24]  
colors $>$ 3.5,  consistent with optically thick 24 $\mu$m  emission (i.e. no object  in 
the sample has a massive disk and  optically thin 24 $\mu$m emission).  
This result excludes any scenario in which most
disks evolve exclusively from the inside out (e.g., an evolution 
dominated by grain growth).   If that were the case, 
we would expect to see some massive disks
with optically thin 24 $\mu$m emission.  Since the 24 $\mu$m 
data probes the inner $\sim$10 AU of the disk, while the vast majority of
the disk mass is outside this radius, any disk evolving strictly 
from the inside out would still retain most of its mass at the point where
its 24 $\mu$m emission is transitioning from optically thick to optically thin. 

Fig. 2 suggests  a scenario in which 
the levels of mid-IR excess emission remain constant while 
the mass of the disk is being depleted by 2 orders of magnitude
through accretion onto the star.  Such a  scenario is illustrated by the 
SEDs in Fig 3. We note that the properties of the SEDs on the bottom of Fig. 3 
and  the top of Fig. 1 overlap. Together, they form a sequence in which a disk 
loses mass,  maintaining a constant near- and mid-IR SED (Fig 3) until the mass 
of the disk reaches a critical level around 1 M$_{Jup}$, at which point the inner 
disk starts to dissipate from  the inside out  through photoevaporation (Fig 1).
The SEDs in Figs. 1 and 3 are ordered,  left to right and top to bottom, 
to \emph{illustrate} this evolutionary sequence.  However,  their order should 
not be taken literately. Other factors,  in addition to the evolutionary status of the disk,  
play a role in the morphology of an SED (e.g.,  disk inclination and stellar luminosity).

\subsection{Implications for Disk Evolution}

As discussed above,  the data in Fig.  2 are inconsistent with
disk evolution occurring strictly from the inside out as expected 
from disk evolution models dominated by grain growth 
(Dullemond $\&$ Dominik, 2005).   
Planet formation is a considerably more complex process
than grain growth. As such, its effect on the observable 
properties of disks are  more uncertain. However, 
planet-formation models still make some testable predictions. In particular,  
forming planets with masses larger than $\sim$0.5-1 M$_{Jup}$ 
should be able to open a gap in the disk independently of the  
disk mass   (Edgar et al.  2007).  The low masses of  \emph{all} 
the disks  in our SMA sample  suggests that  their 
inner holes are not driven mainly by the formation of Jovian 
planets.  However, the fact that the vast majority of our targets are
non-accreting objects could introduce a strong bias against 
disks with planet-induced inner holes because, unless the 
planet is very massive ($>$10 M$_{Jup}$), some disk material 
is expected to flow  across the planet's orbit and reach the star
(Lubow et al. 1999). 
%
The SEDs of our sample are most consistent with those of photoevaporating disks.
Photoevaporation models, such as those presented by 
Alexander et al. (2006), can simultaneously explain  not only  the 
properties of our SMA sample, but also many of the observational 
results discussed in this Letter, most notably:

{ \it{a) The duration of the CTTS stage, the fact that most of their IR SEDs 
look alike,  and their wide range of  disk masses.}}  According to  
photoevaporation models,  during the first few Mys of  evolution 
viscous accretions dominate over photoevaporation.   
The  mass of the  disk is depleted by accretion onto the star,  but   the accretion rate 
through the disk is large enough to replenish the inner disk.  Thus,  the 
IR SED remains unaffected,  as seen in Fig 3.

{\it{b) The fraction of WTTS with disks, their SEDs, and their low 
disk masses.}}  Photoevaporation  predicts that all disks pass 
through a short ($\tau$ $<$ 0.5 Mys) inside-out clearing phase \
once the accretion rate matches the photoevaporation rate  ($\sim$10$^{-10}$ 
M$_{\odot}$yr$^{-1}$).  During this clearing stage, the mass of the disk is 
predicted to be 0.05-0.5 M$_{Jup}$, depending on the exact 
ionizing  flux and the viscosity law.   This phase is in excellent 
agreement with the properties and incidence of
WTTS disks (See Fig. 1).  Also, we note  that FW Tau, the transition 
object highlighted in Fig. 2, has a disk with a mass of 
$\sim$0.4 M$_{Jup}$, within the  predicted range. 

{\it{c) The low incidence of  holes in disks that are massive or strongly accreting.}}  
Photoevaporation predicts that the inner disk will drain \emph{only} after the 
outer disk has been significantly depleted of mass  and the accretion rate 
becomes very small.  Counter examples to this prediction exist, such as
GM Aur and DM Tau (M$_{DISK}$ $\sim$25 M$_{Jup}$, 
M$_{ACC}$= $\sim$10$^{-8}$ 
M$_{\odot}$   yr$^{-1}$, Najita et al. 2007), but their incidence seems to be of the order of a 
few percent. Other processes, such as grain growth or planet formation, must be 
responsible for their inner holes.  

In practice, all the processes discussed in this Letter (grain growth, 
planet formation, and photoevaporation) are expected to operate 
simultaneously and affect one another.  Other processes
such as dynamical interactions in binary stellar systems  are also likely to play a 
role (Ireland $\&$ Kraus, 2008).  However, the remarkable success of the 
photoevaporation models accounting for the many observational results 
listed above strongly suggests that,  together with viscous accretion, 
photoevaporation is one of the dominant processes driving disk evolution.
This conclusion seems to contradict the recent results by Najita et al. (2007),
who find that only 2 out of the 12 transition objects they consider are
consistent with photoevaporating disks. However, the different sample
selection biases likely account for the discrepancy (Alexander 2008).
Quantifying the relative importance of disk evolution mechanisms requires 
establishing specific observational metrics to distinguish among them and taking 
into consideration the details of  the sample selection.  Such a task is 
beyond the scope of this Letter, but will be attempted in a followup paper.  
 
\acknowledgments
We thanks Richard Alexander, Nairn Baliber, and the anonymous referee for 
their valuable comments.  Support for this work was provided by NASA through 
the \emph{Spitzer} Fellowship Program. 
J.P.W. $\&$ G.S.M. acknowledge support from NSF grant AST08-08144. 
This work makes use of  data obtained by the \emph{Spitzer} Space Telescope
and the 2MASS survey,  which is a joint project of the University of 
Massachusetts and IPAC/Caltech.


\begin{deluxetable}{lcrccr}
\tablewidth{0pt}
\tablecaption{SMA Sample}
\tablehead{
\colhead{Name}&\colhead{K$_S$}&\colhead{24 $\mu$m\tablenotemark{1}}&
\colhead{850 $\mu$m\tablenotemark{2}}   &\colhead{1300 $\mu$m\tablenotemark{2}}  &\colhead{M$_{disk}$}  \\
\colhead{ }&\colhead{(mag)} &\colhead{(mJy)} &\colhead{(mJy)}&
\colhead{(mJy)}&\colhead{(M$_{Jup}$)} 
}
\startdata
TWA 7    &     6.90 &       30    &   $<$16          & \nodata     & $<$0.2 \\
TWA 3    &     6.77 &   1650    &   \nodata       &  47$\pm$3&       1.0  \\
TWA 13  &     7.50 &        18   &   $<$16         & \nodata      & $<$0.2 \\
TWA 4    &      5.58 &    8500  &   115$\pm$5& 64$\pm$4 &       1.2  \\
TWA 11  &     5.77 &   3030    &   \nodata        & $<$9.4      & $<$0.2\\
\enddata
\tablecomments{This table is only a sample of the content. The complete version is in the electronic edition of
the Journal.  
}
\tablenotetext{1}{
TWA  and Oph  data from  Cieza et al. (2007) and 
Low et al. (2005), respectively. Upper Sco photometry obtained 
from archival data using the c2d pipeline (Evans et al. 2007).}
\tablenotetext{2}{
Errors are 1-$\sigma$ statistical uncertainties. Limits are 3-$\sigma$.}\vspace{0.5mm}
\end{deluxetable}


\begin{figure}
\epsscale{0.7}
\plotone{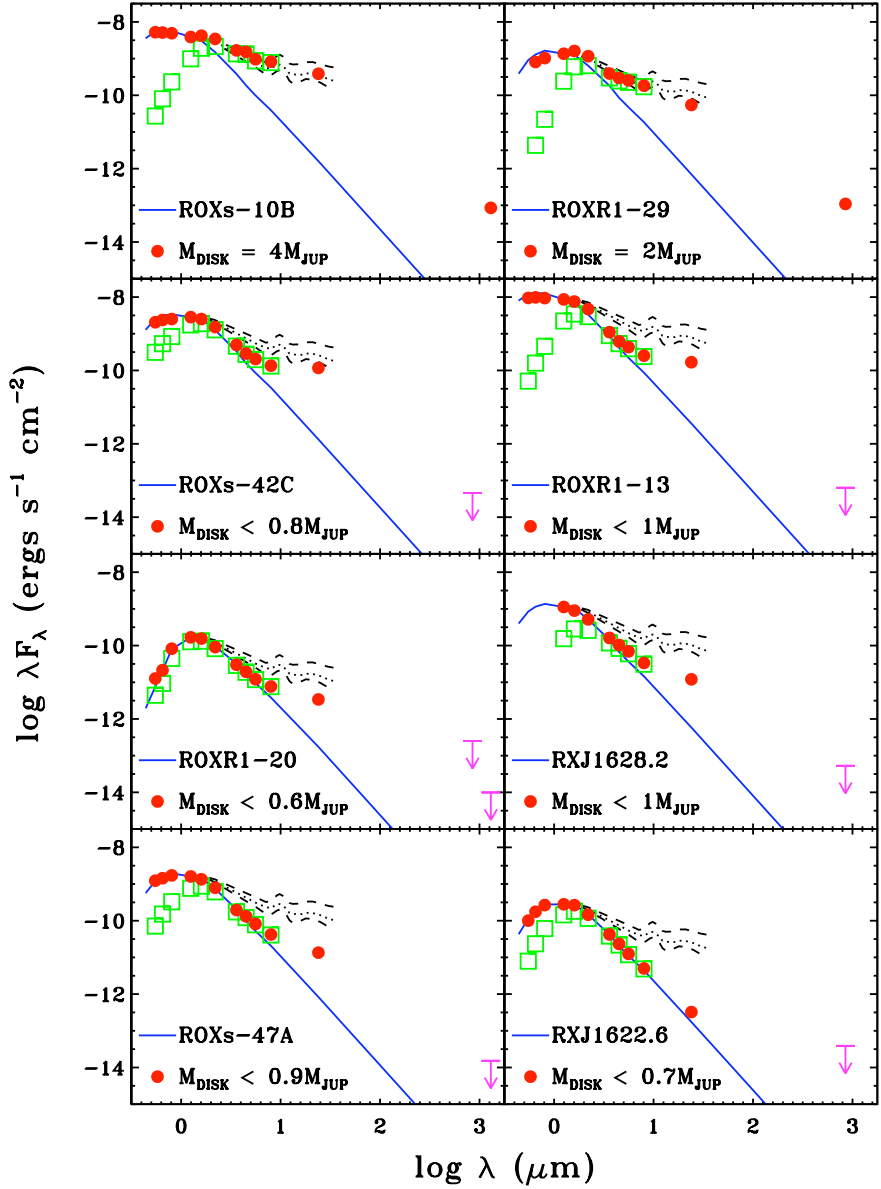}
\caption{
The optical to (sub)millimeter SEDs of 
some of our SMA targets. The filled circles are detections 
while the arrows represent 3-$\sigma$ upper limits.  The open squares correspond 
to the observed optical and near-IR fluxes before being corrected for extinction as described by Cieza et al. (2007). 
The dotted lines correspond to the median SED of K5-M2 CTTSs calculated by Furlan et al. (2006).
The dashed lines are the quartiles. These SEDs trace the inside-out dissipation of the inner disk.
}
\end{figure}

\begin{figure}
\epsscale{1.0}
\plotone{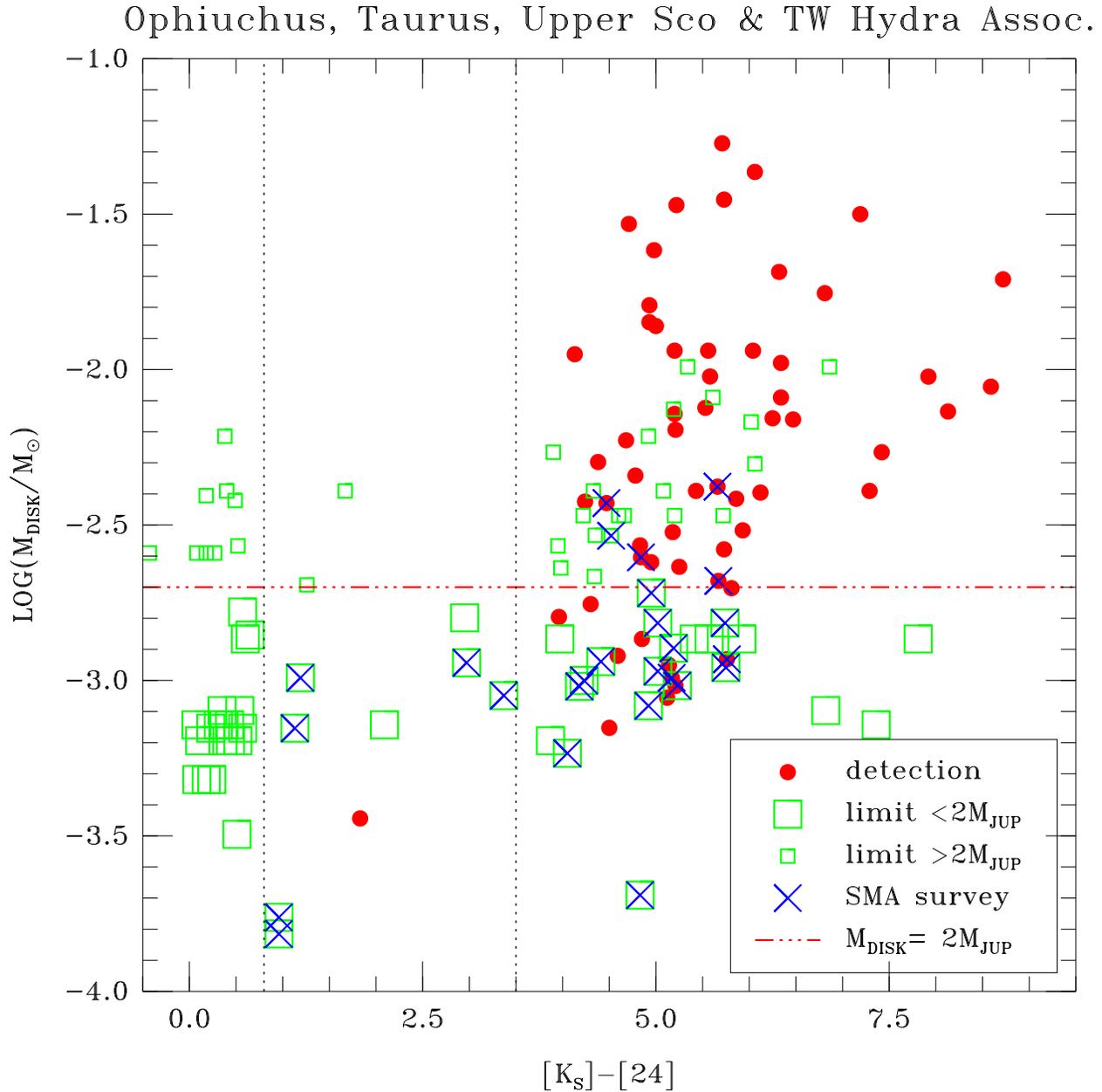}
\caption{
PMS disk masses as a function of [K$_S$]-[24]  color for our extended sample
The symbols  are as labeled in the figure.  Our SMA 3-$\sigma$ upper limits are all 
$\lesssim$ 2 M$_{Jup}$. Higher, less significant  limits are shown as smaller symbols.  
The vertical dotted lines divide the sample into three groups (left to right): 
bare stellar photospheres, 
disks likely to be optically thin at  24 $\mu$m, and  
disks likely to be optically thick at  24 $\mu$m.
The only (sub)millimeter detection outside the optically thick group is FW Tau 
with a disk mass of $\sim$0.4 M$_{Jup}$.
}
\end{figure}

\begin{figure}
\figurenum{3}
\epsscale{0.7}
\plotone{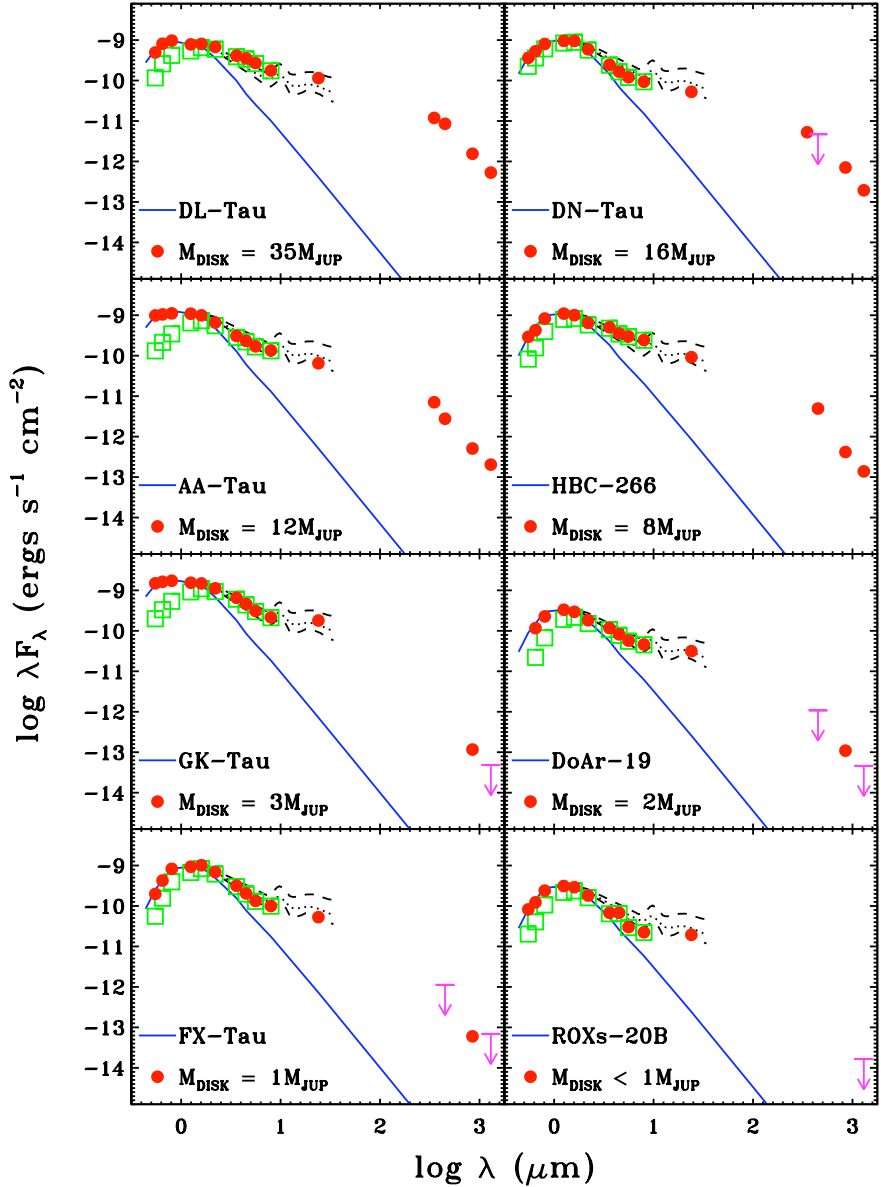}
\caption{
The SEDs of some of the objects from Figure 2 that have [K]-[24]  $\sim$5
but widely different disk masses.The symbols are the same as in Fig. 1. 
We argue that these objects represent an evolutionary sequence
\emph{prior} to that  illustrated by Fig. 1. Here, the mass of the disk is being depleted by 2 orders of 
magnitude through accretion onto the star. The near- and mid-IR SEDs remain unaffected until the 
disk mass reaches the 1 M$_{Jup}$ level, at which point the disk drains from the inside out, as traced 
by the SEDs in Figure 1. 
\vspace{0.5mm}
}
\end{figure}

\end{document}